\documentclass[aps,twocolumn,showpacs]{revtex4}
\usepackage{amsmath,amssymb}
\usepackage{graphics}
\usepackage{dcolumn,bm}

\begin{document}
\title{Kinetic market models with single commodity having price fluctuations}

\author{
Arnab Chatterjee
}
%\thanks{\emph{Email address:} 
\email{arnab.chatterjee@saha.ac.in}
\author{
%\and
Bikas K. Chakrabarti
%\thanks{\emph{Email address:} 
}
\email{bikask.chakrabarti@saha.ac.in}

\affiliation{Theoretical Condensed Matter Physics Division and 
Centre for Applied Mathematics and Computational Science,\\ 
Saha Institute of Nuclear Physics, 1/AF Bidhannagar, Kolkata 700064, India.}

%%%%%%%%%%%%%%%%%%%%%%%%%%%%%%%%%%%%%%%%%%%%%%%%%%%%%%%%%%%%%%%%
\begin{abstract}
%\abstract{
We study here numerically the behavior of an ideal gas like model of markets having
only one non-consumable commodity. We investigate the behavior of the 
steady-state distributions of money, commodity and total wealth,
as the dynamics of trading or exchange of money and commodity proceeds,
with local (in time) fluctuations in the price of the commodity.
These distributions are studied in markets with agents having uniform and 
random saving factors. The self-organizing features in money distribution 
are similar to the cases without any commodity (or with consumable 
commodities), while the commodity distribution shows an exponential decay. 
The wealth distribution shows interesting behavior: Gamma like
distribution for uniform saving propensity and has the same power-law tail,
as that of the money distribution, for 
a market with agents having random saving propensity.
\end{abstract}

\maketitle
%%%%%%%%%%%%%%%%%%%%%%%%%%%%%%%%%%%%%%%%%%%%%%%%%%%%%%%%%%%%%%%%
%%%%%%%%%%%%%%%%%%%%%%%%%%%%%%%%%%%%%%%%%%%%%%%%%%%%%%%%%%%%%%
\section{Introduction}
%%%%%%%%%%%%%%%%%%%%%%%%%%%%%%%%%%%%%%%%%%%%%%%%%%%%%%%%%%%%%%
%%%%%%%%%%%%%%%%%%%%%%%%%%%%%%%%%%%%%%%%%%%%%%%%%%%%%%%%%%%%%%%%
The study of wealth distribution~\cite{cc:EWD05} in a society has 
remained an intriguing problem since Vilfredo Pareto who first
observed~\cite{cc:Pareto:1897} that the number of rich people with 
wealth $m$ decay following an inverse power-law:
\begin{equation}
P(m) \sim m^{-(1+\nu)}.
\label{par}
\end{equation}
$P(m)$ is the number density of people possessing wealth $m$
and $\nu$ is known as the Pareto exponent. This exponent generally
assumes a value between $1$ and $3$ in different economies and 
times~\cite{cc:EWD05,cc:realdatag,cc:realdataln,cc:Sinha:2006}.
It is also known that for low and medium income, the number density
$P(m)$ falls off much faster: exponentially~\cite{cc:realdatag} or
in a log-normal way~\cite{cc:realdataln}.

In recent years, easy availability of data in electronic media has helped in the
analysis of wealth or income distributions in various societies~\cite{cc:EWD05}.
It is now more or less established that the distribution
has a power-law tail for the large (about 5\% of the population)
wealth/income
while the majority (around 95\%) low income distribution fits well to
Gibbs or log-normal 
form~\cite{cc:EWD05,cc:Pareto:1897,cc:realdatag,cc:realdataln,cc:Sinha:2006,cc:othermodels}.

There have been several attempts to model a simple economy
with minimum trading ingredients, 
which involve a wealth exchange process~\cite{cc:EWD05}
that produce a distribution
of wealth similar to that observed in the real market.
We  are particularly interested in microscopic models of markets where
the (economic) trading activity is considered as a scattering
process~\cite{cc:marjit,cc:Dragulescu:2000,cc:Chakraborti:2000,cc:Hayes:2002,cc:Chatterjee:2004,cc:Chatterjee:2003,cc:Chakrabarti:2004,cc:Slanina:2004}
(see also Ref.~\cite{cc:ESTP:KG} for recent extensive reviews).
We concentrate on models that incorporate `saving propensity' (of the agents) 
as an essential ingredient in a trading process, and reproduces the salient 
features seen across wealth distributions in varied economies
(see Ref.~\cite{cc:EWD:CC} for a review).
Much earlier, Angle~\cite{cc:Angle:1986} studied an inequality process, which
can be mapped to the uniform savings models is certain 
cases; see Ref.~\cite{cc:Angle:2006} for a detailed review.

These studies also show (and discussed briefly here) 
how the distribution of savings can
be modified to reproduce the salient features of empirical
distributions of wealth -- namely the shape of the distribution
for the low and middle wealth and the tunable Pareto exponent.
In all these 
models~\cite{cc:Chakraborti:2000,cc:Hayes:2002,cc:Chatterjee:2004,cc:Chatterjee:2003,cc:Chakrabarti:2004},
`savings' was introduced as a quenched parameter that
remained invariant with time (or trading). 

Apart from presenting a brief summary in Section II (giving the 
established results in such models), 
we present new results for a similar (gas like) market model,
where the exchange is for a non-consumable commodity (globally conserved,
like money). We find, although the details of the steady-state money
and wealth (money and commodity together) distributions differ
considerably, the same Pareto tail feature appears in both, with identical
exponent ($\nu$) value.

%%%%%%%%%%%%%%%%%%%%%%%%%%%%%%%%%%%%%%%%%%%%%%%%%%%%%%%%%%%%%%%%
%%%%%%%%%%%%%%%%%%%%%%%%%%%%%%%%%%%%%%%%%%%%%%%%%%%%%%%%%%%%%%
\section{Ideal-gas like models of trading markets without any commodity}
%%%%%%%%%%%%%%%%%%%%%%%%%%%%%%%%%%%%%%%%%%%%%%%%%%%%%%%%%%%%%%
%%%%%%%%%%%%%%%%%%%%%%%%%%%%%%%%%%%%%%%%%%%%%%%%%%%%%%%%%%%%%%%%
%%%%%%%%%%%%%%%%%%%%%%%%%%%%%%%%%%%%%%%%%%%%%%%%%%%%%%%%%%%%%%
\subsection{Without savings}
%%%%%%%%%%%%%%%%%%%%%%%%%%%%%%%%%%%%%%%%%%%%%%%%%%%%%%%%%%%%%%
We first consider an ideal-gas model of a closed economic system.
Wealth is measured in terms of the amount of money possessed by an
individual.
Production is not allowed i.e, total money $M$ is fixed and  also there
is no migration of population i.e, total number of agents $N$ is fixed, 
and the only economic activity is confined to trading.
Each agent $i$, individual or corporate, possess money $m_i(t)$ at time $t$.
In any trading, a pair of agents $i$ and $j$ randomly exchange their 
money~\cite{cc:marjit,cc:Dragulescu:2000,cc:Chakraborti:2000}, 
such that their total money is (locally) conserved
and none posses negative money ($m_i(t) \ge 0$, i.e, debt not allowed):
\begin{equation} 
\label{mdelm}
m_i(t+1) = m_i(t) + \Delta m; \  m_j(t+1) = m_j(t) - \Delta m \nonumber \\
\end{equation} 
\begin{equation} 
\label{deltam}
\Delta m = \epsilon (m_i(t)+m_j(t)) - m_i(t); \ 0 \le \epsilon \le 1
\end{equation} 
All the money transactions considered in this paper follow local
conservation:
\begin{equation} 
\label{consv}
m_i(t) + m_j(t) = m_i(t+1) + m_j(t+1).
\end{equation}
The time ($t$) changes by one unit after each trading and $\epsilon$
is a random fraction chosen independently for each trading
or at each time $t$.
The steady-state ($t \to \infty$) distribution of money is Gibbs one:
\begin{equation}
\label{gibbs}
P(m)=(1/T)\exp(-m/T);T=M/N. 
\end{equation} 
No matter how uniform or justified the initial distribution is, the
eventual steady state corresponds to Gibbs distribution where most of the
people have very little money.
This follows from the conservation of money and additivity of entropy:
\begin{equation}
\label{prob}
P(m_1)P(m_2)=P(m_1+m_2) \nonumber.
\end{equation}
This steady state result is quite robust and realistic.
Several variations of the trading~\cite{cc:EWD05}, 
does not affect the distribution. 
%, and of the `lattice'
%(on which the agents can be put and each agent trade with its
%`lattice neighbors' only) --- compact, fractal or small-world 
%like

In any trading, savings come naturally~\cite{cc:Samuelson:1980}.
A saving factor $\lambda$ is therefore introduced in the same 
model~\cite{cc:Chakraborti:2000} (Ref.~\cite{cc:Dragulescu:2000} 
is the model without savings), where each trader
at time $t$ saves a fraction $\lambda$ of its money $m_i(t)$ and trades
randomly with the rest.
In each of the following two cases, the savings fraction does
not vary with time, and hence we call it `quenched' in the terminology
of statistical mechanics.

%%%%%%%%%%%%%%%%%%%%%%%%%%%%%%%%%%%%%%%%%%%%%%%%%%%%%%%%%%%%%%
\subsection{Uniform savings}
%%%%%%%%%%%%%%%%%%%%%%%%%%%%%%%%%%%%%%%%%%%%%%%%%%%%%%%%%%%%%%
For the case of `uniform' savings, the money exchange rules
remain the same (Eqn.~(\ref{mdelm})), where
\begin{equation}
\label{eps}
\Delta m=(1-\lambda )[\epsilon \{m_{i}(t)+m_{j}(t)\}-m_{i}(t)],
\end{equation}
where $\epsilon$ is a random fraction, coming from the stochastic nature
of the trading. $\lambda$ is a fraction ($0 \le \lambda < 1$)
which we call the saving factor.

The market (non-interacting at $\lambda =0$ and $1$) becomes effectively
`interacting'
for any non-vanishing $\lambda \; (<1)$: For uniform $\lambda$ (same for all
agents), the steady state distribution $P_f(m)$ of money is sharply
decaying on both sides with the most-probable money per agent shifting away
from $m=0$ (for $\lambda =0$) to $M/N$ as $\lambda \to 1$.
The self-organizing feature of this market,
induced by sheer self-interest of saving by each agent without any global
perspective, is very significant as the fraction of paupers decrease with
saving fraction $\lambda$ and most people possess some fraction of the
average money in the market (for $\lambda \to 1$, the socialists'
dream is achieved with just people's self-interest of saving!).
This uniform saving propensity does not give the Pareto-like
power-law distribution yet, 
but the Markovian nature of the scattering or trading
processes (eqn.~(\ref{prob})) is lost and the system becomes co-operative.
Through $\lambda$, the agents indirectly get to develop a correlation with
(start interacting with) each other and the system co-operatively 
self-organizes~\cite{cc:Bak:1997} towards a most-probable distribution.

This model has been understood to a certain extent 
(see e.g,~\cite{cc:Das:2003,cc:Patriarca:2004,cc:Repetowicz:2005}), 
and argued to resemble a gamma distribution~\cite{cc:Patriarca:2004}, 
and partly explained analytically.
This model clearly finds its relevance in cases where the economy consists
of traders with `waged' income~\cite{cc:Willis:2004}.

%%%%%%%%%%%%%%%%%%%%%%%%%%%%%%%%%%%%%%%%%%%%%%%%%%%%%%%%%%%%%%
\subsection{Distributed savings}
%%%%%%%%%%%%%%%%%%%%%%%%%%%%%%%%%%%%%%%%%%%%%%%%%%%%%%%%%%%%%%
In a real society or economy, $\lambda$ is a very inhomogeneous parameter:
the interest of saving varies from person to person.
We move a step closer to the real situation where saving factor $\lambda$ is
widely distributed within the 
population~\cite{cc:Chatterjee:2004,cc:Chatterjee:2003,cc:Chakrabarti:2004}.
The evolution of money in such a trading
can be written as:
\begin{eqnarray}
\label{mi}
m_i(t+1)&=&\lambda_i m_i(t) + \epsilon \nonumber\\ 
&&  \times \left[(1-\lambda_i)m_i(t) + (1-\lambda_j)m_j(t)\right], 
\end{eqnarray}
\begin{eqnarray}
\label{mj}
m_j(t+1)&=&\lambda_j m_j(t) + (1-\epsilon) \nonumber\\
&&  \times \left[(1-\lambda_i)m_i(t) + (1-\lambda_j)m_j(t)\right]
\end{eqnarray}
One again follows the same rules (Eqn.~(\ref{mdelm})) as before, except that
\begin{equation}
\label{lrand}
\Delta m=(1-\lambda_{j})\epsilon m_{j}(t)-(1-\lambda _{i})(1 - \epsilon)m_{i}(t)
\end{equation}
here; $\lambda _{i}$ and $\lambda _{j}$ being the saving
propensities of agents $i$ and $j$. The agents have uniform (over time) saving
propensities, distributed independently, randomly and uniformly (white)
within an interval $0$ to $1$ agent $i$ saves a random fraction
$\lambda_i$ ($0 \le \lambda_i < 1$) and this $\lambda_i$ value is quenched
for each agent ($\lambda_i$ are independent of trading or $t$).
$P(m)$ is found to follow a strict power-law decay.
This decay fits to Pareto
law (\ref{par}) with $\nu = 1.01 \pm 0.02$ for several decades.
This power law is extremely robust: a distribution 
\begin{equation}
\label{lam0}
\rho(\lambda) \sim |\lambda_0-\lambda|^\alpha, 
\ \ \lambda_0 \ne 1, \ \ 0 \le \lambda<1,
\end{equation}
of quenched $\lambda$ values among the agents produce power law
distributed $m$ with Pareto index $\nu=1$, irrespective of the value of
$\alpha$. For negative $\alpha$ values, however,
we get an initial (small $m$) Gibbs-like decay in $P(m)$.
In case $\lambda_0 =1$, the Pareto exponent is modified to
$\nu=1+\alpha$, which qualifies for the non-universal exponents
in the same model \cite{cc:Chatterjee:2004,cc:Mohanty:2006}.

This model~\cite{cc:Chatterjee:2004} has been thoroughly analyzed, 
and the analytical
derivation of the Pareto exponent has been achieved in certain 
cases~\cite{cc:Repetowicz:2005,cc:Mohanty:2006,cc:Chatterjee:2005}.
The Pareto exponent has been derived to exactly $1$.

In this model, agents with higher saving propensity tend to hold 
higher average wealth, which is justified by the fact that the saving
propensity in the rich population is always high~\cite{cc:Dynan:2004}.

%%%%%%%%%%%%%%%%%%%%%%%%%%%%%%%%%%%%%%%%%%%%%%%%%%%%%%%%%%%%%%
%%%%%%%%%%%%%%%%%%%%%%%%%%%%%%%%%%%%%%%%%%%%%%%%%%%%%%%%%%%%%%
\section{Ideal-gas trading market in presence of a non-consumable commodity}
%%%%%%%%%%%%%%%%%%%%%%%%%%%%%%%%%%%%%%%%%%%%%%%%%%%%%%%%%%%%%%
%%%%%%%%%%%%%%%%%%%%%%%%%%%%%%%%%%%%%%%%%%%%%%%%%%%%%%%%%%%%%%
In the above markets, modifications due to exchange of a consumable 
commodity hardy affects the distribution, as the commodity once bought 
or sold need not be accounted for. 
Consumable commodities effectively have no `price', as due to their short
lifetime to contribute to the total wealth of an individual.
It is interesting however, to study the role of non-consumable commodities
in such market models and this we do here.

In the simplified version of a market with a single non-consumable
commodity, we again consider a fixed number of traders or agents $N$
who trade in a market involving total money $\sum_i m_i(t)=M$ and total
commodity $\sum_i c_i(t)=C$, $m_i(t)$ and $c_i(t)$ being the money and commodity
of the $i$-th agent at time $t$ and are both non-negative. 
Needless to mention, both $m_i(t)$ and $c_i(t)$ change with time or trading
$t$. 
The market, as before is closed, which means, $N$, $M$ and $C$
are constants. 
The wealth $w_i$ of an individual $i$ is thus, the sum of the money
and commodity it possesses, i.e., $w_i=m_i + p_0 c_i$; 
$p_0$ is the ``global'' price.
In course of trading, total money and total commodity are locally conserved,
and hence the total wealth.
In such a market, one can define a global average price parameter 
$p_0=M/C$, which is set here to unity, giving $w_i = m_i + c_i$. 
It may be noted here that in order to avoid the complication of restricting
the commodity-money exchange and their reversal between the same agents,
the Fisher velocity of money circulation (see e.g., Ref.~\cite{cc:Wang:ESTP})
is renormalised to unity here.
In order to accommodate the lack 
of proper information and the ability of the agents to bargain etc., we will
allow of course fluctuations $\delta$ in the price of the commodities
at any trading (time): $p(t)= p_0 \pm \delta = 1 \pm \delta$.
We find, the nature of steady state to be unchanged and independent of
$\delta$, once it becomes nonvanishing.

%%%%%%%%%%%%%%%%%%%%%%%%%%%%%%%%%%%%%%%%%%%%%%%%%%%%%%%%%%%%%%%%5
\begin{figure}%[t]
\resizebox{8.5cm}{!}{
\includegraphics{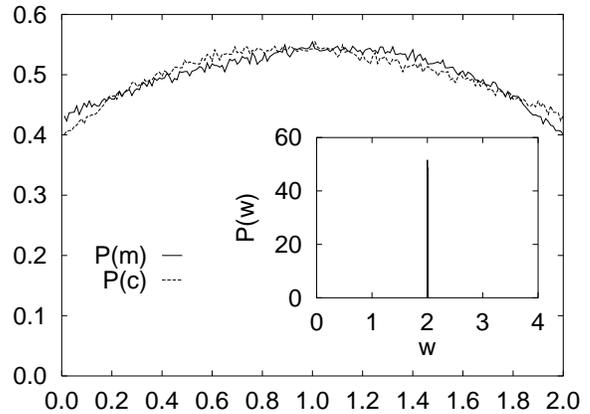}
}
\caption{
Steady state distribution $P(m)$ of money $m$ in a market with no savings
(saving factor $\lambda=0$) for no price fluctuations i.e, $\delta=0$.
The graphs show simulation results for a system of $N=100$ agents, 
$M/N=1$, $C/N=1$; $m_i=1=c_i$ at $t=0$ for all agents $i$.
The inset shows the distribution $P(w)$ of total wealth $w=m+c$.
As $p=1$, for $\delta=0$, although $m$ and $c$ can change with tradings
within the limit $(0-2)$ the sum is always maintained at $2$.
}
\label{fig:commogif.flam.del0}
\end{figure}
%%%%%%%%%%%%%%%%%%%%%%%%%%%%%%%%%%%%%%%%%%%%%%%%%%%%%%%%%%%%%%%%5
%%%%%%%%%%%%%%%%%%%%%%%%%%%%%%%%%%%%%%%%%%%%%%%%%%%%%%%%%%%%%%%%5
\subsection{Dynamics}
%%%%%%%%%%%%%%%%%%%%%%%%%%%%%%%%%%%%%%%%%%%%%%%%%%%%%%%%%%%%%%%%5
In general, the dynamics of money in this market looks the same as
Eqn.~(\ref{mdelm}), with $\Delta m$ given by
Eqns.~(\ref{deltam}), (\ref{eps}) or (\ref{lrand}) depending on whether
$\lambda_i=0$ for all, $\lambda_i \ne 0$ but uniform for all $i$ or
$\lambda_i \ne \lambda_j$ respectively.
However, all $\Delta m$ are not allowed here; only those, for which
$\Delta m_i \equiv m_i(t+1)-m_i(t)$ or $\Delta m_j$ are allowed by the
corresponding changes $\Delta c_i$ or $\Delta c_j$ 
in their respective commodities ($\Delta m > 0, \Delta c > 0$):
\begin{equation}
\label{commo-eqn1}
c_i(t+1)=c_i(t)+ \frac{m_i(t+1)-m_i(t)}{p(t)}
\end{equation}
\begin{equation}
\label{commo-eqn2}
c_j(t+1)=c_j(t)- \frac{m_j(t+1)-m_j(t)}{p(t)}
\end{equation}
where $p(t)$ is the local-time `price' parameter, a stochastic variable:
\begin{equation}
\label{pricedelta}
p(t)=
\left\{\begin{array}{c}
1 + \delta {\rm \;\; with \; probability \;0.5}\\
1 - \delta {\rm \;\; with \; probability \;0.5}
\end{array}\right..
\end{equation}
The role of the stochasticity in $p(t)$ is to imitate the effect of
bargaining in a trading process. $\delta$ parametrizes the amount of
stochasticity. The role of $\delta$ is significant in the sense that it 
determines the (relaxation) time the whole system takes to reach a dynamically
equilibrium state; the system reaches equilibrium sooner for larger
$\delta$, while its magnitude does not affect the steady state distribution.
It may be noted that, in course of trading process, certain exchanges
are not allowed (e.g., in cases when a particular pair of traders do not
have enough commodity to exchange in favor of an agreed exchange of money).
We then skip these steps and choose a new pair of agents for trading.
%%%%%%%%%%%%%%%%%%%%%%%%%%%%%%%%%%%%%%%%%%%%%%%%%%%%%%%%%%%%%%%%5
\begin{figure}%[t]
\resizebox{8.5cm}{!}{
\includegraphics{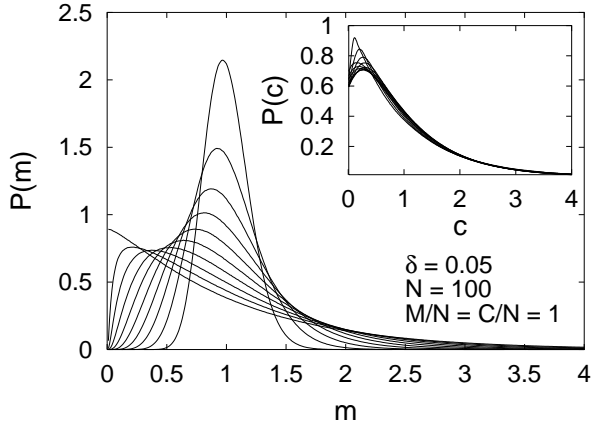}
}
\caption{
Steady state distribution $P(m)$ of money $m$ in the 
uniform savings commodity market 
for different values of saving factor $\lambda$ 
($0,0.1,0.2,0.3,0.4,0.5,0.6,0.7,0.8,0.9$ from left to right near the origin)
for $\delta=0.05$. 
The inset shows the distribution $P(c)$ of commodity $c$ in the uniform 
savings commodity market for different values of saving factor $\lambda$.
The graphs show simulation results for a system of
$N=100$ agents, $M/N=1$, $C/N=1$.
}
\label{fig:commogif.flam.mc}
\end{figure}
%%%%%%%%%%%%%%%%%%%%%%%%%%%%%%%%%%%%%%%%%%%%%%%%%%%%%%%%%%%%%%%%5
%%%%%%%%%%%%%%%%%%%%%%%%%%%%%%%%%%%%%%%%%%%%%%%%%%%%%%%%%%%%%%%%5
\begin{figure}%[t]
\resizebox{8.5cm}{!}{
\includegraphics{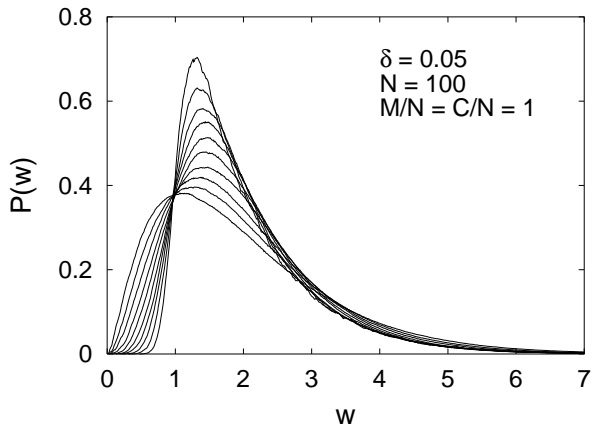}
}
\caption{
Steady state distribution $P(w)$ of total wealth $w=m+c$ 
in the uniform savings commodity 
market for different values of saving factor $\lambda$ 
($0,0.1,0.2,0.3,0.4,0.5,0.6,0.7,0.8,0.9$ from left to right)
for $\delta=0.05$.  The graphs show simulation results for a system of
$N=100$ agents, $M/N=1$, $C/N=1$.
}
\label{fig:commogif.flam.w}
\end{figure}
%%%%%%%%%%%%%%%%%%%%%%%%%%%%%%%%%%%%%%%%%%%%%%%%%%%%%%%%%%%%%%%%5

%%%%%%%%%%%%%%%%%%%%%%%%%%%%%%%%%%%%%%%%%%%%%%%%%%%%%%%%%%%%%%%%5
\subsection{Results}
%%%%%%%%%%%%%%%%%%%%%%%%%%%%%%%%%%%%%%%%%%%%%%%%%%%%%%%%%%%%%%%%5
For $\delta=0$, of course, the wealth of each agent remains invariant 
with time; only the proportion of money and commodity interchange 
within themselves, since the `price' factor remains constant.
This of course happens irrespective of the savings factor being zero,
uniform or distributed. For $\delta=0$, the steady state distribution
of money or commodity can take non-trivial forms: 
(see Fig.~\ref{fig:commogif.flam.del0}), but has strictly a 
$\delta$-function behavior for the total wealth distribution; 
it gets frozen at the value of wealth one starts with (see inset of 
Fig.~\ref{fig:commogif.flam.del0} for the case $m_i=1=c_i$ for all $i$).

%%%%%%%%%%%%%%%%%%%%%%%%%%%%%%%%%%%%%%%%%%%%%%%%%%%%%%%%%%%%%%%%5
\begin{figure}%[t]
\resizebox{8.5cm}{!}{
\includegraphics{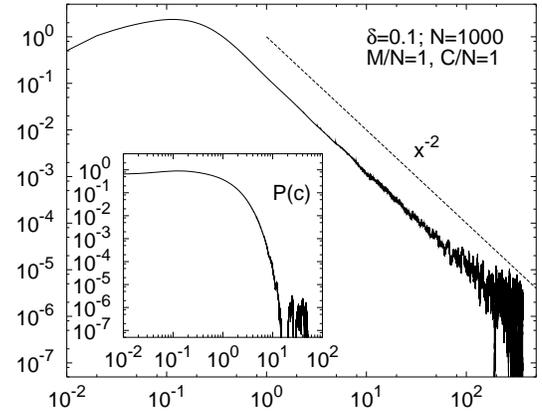}
}
\caption{
Steady state distribution $P(m)$ of money $m$ 
in the commodity market with distributed savings $0 \le \lambda < 1$.
$P(m)$ has a power-law tail with Pareto index $\nu=1 \pm 0.02$
(a power law function $x^{-2}$ is given for comparison).
The inset shows the distribution $P(c)$ of commodity $c$
in the same commodity market.
The graphs show simulation results for a system of
$N=1000$ agents, $M/N=1$, $C/N=1$.
}
\label{fig:commogif.dlam.mc}
\end{figure}
%%%%%%%%%%%%%%%%%%%%%%%%%%%%%%%%%%%%%%%%%%%%%%%%%%%%%%%%%%%%%%%%5
As mentioned already for $\delta \ne 0$, the steady state results
are not dependent on the value of $\delta$ (the relaxation time
of course decreases with increasing $\delta$).
In such a market with uniform savings, money distribution $P(m)$ has a form
similar to a set (for $\lambda \ne 0$) 
of Gamma functions (see Fig.~\ref{fig:commogif.flam.mc}):
a set of curves with a most-probable value shifting from $0$ to $1$
as saving factor $\lambda$ changes from  $0$ to $1$
(as in the case without any commodity).
The commodity distribution $P(c)$ has an initial peak and an exponential
fall-off, without much systematics with varying 
$\lambda$ (see inset of Fig.~\ref{fig:commogif.flam.mc}).
The distribution $P(w)$ of total wealth $w=m+c$ behaves much like $P(m)$
(see Fig.~\ref{fig:commogif.flam.w}). 
It is to be noted that since there is no precise correspondence with
commodity and money for $\delta \ne 0$
(unlike when $\delta=0$, when the sum is fixed),
$P(w)$ cannot be derived directly from $P(m)$ and $P(c)$.
However, there are further interesting
features. Although they form a class of Gamma distributions, the set of
curves for different values of saving factor $\lambda$ seem to intersect
at a common point, near $w=1$. All the reported data are for a system
of $N=100$ agents, with $M/N=1$ and $C/N=1$ and for a case where the noise 
level $\delta$ equals $10$\%.
%%%%%%%%%%%%%%%%%%%%%%%%%%%%%%%%%%%%%%%%%%%%%%%%%%%%%%%%%%%%%%%%5
\begin{figure}%[t]
\resizebox{8.5cm}{!}{
\includegraphics{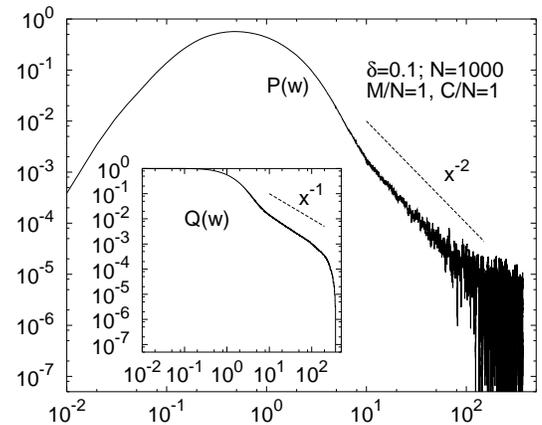}
}
\caption{
Steady state distribution $P(w)$ of total wealth $w=m+c$ 
in the commodity market with distributed savings $0 \le \lambda < 1$.
$P(w)$ has a power-law tail with Pareto index $\nu=1 \pm 0.05$
(a power law function $x^{-1}$ is given for comparison).
The inset shows the cumulative distribution $Q(w) \equiv \int_w^\infty P(w) {\rm d}w$. 
The graphs show simulation results for a system of
$N=1000$ agents, $M/N=1$, $C/N=1$.
}
\label{fig:commogif.dlam.w}
\end{figure}
%%%%%%%%%%%%%%%%%%%%%%%%%%%%%%%%%%%%%%%%%%%%%%%%%%%%%%%%%%%%%%%%5

For $\lambda$ distributed uniformly within the interval $0 \le \lambda < 1$,
the tails of both money and wealth distributions $P(m)$ and $P(w)$ have 
Pareto law behavior with a fitting exponent value
$\nu=1 \pm 0.02$ and $\nu=1 \pm 0.05$ respectively 
(see Figs.~\ref{fig:commogif.dlam.mc} and \ref{fig:commogif.dlam.w} 
respectively),
whereas the commodity distribution is still exponentially decaying
(see inset of Fig.~\ref{fig:commogif.dlam.mc}).

%%%%%%%%%%%%%%%%%%%%%%%%%%%%%%%%%%%%%%%%%%%%%%%%%%%%%%%%%%%%%%%%5
\section{Summary and conclusions}
%%%%%%%%%%%%%%%%%%%%%%%%%%%%%%%%%%%%%%%%%%%%%%%%%%%%%%%%%%%%%%%%5
Let us first summarize the results for the models without any commodity
(money-only exchange models):
There are $N$ players participating in a game,
each having an initial capital of one unit of money. $N$ is very large,
and total money $M=N$ remains fixed over the game (so also the number of
players $N$). 
(a) In the simplest version, the only move at any time is that two 
of these players are
randomly chosen and they decide to share their total money randomly
among them. As one can easily guess, the initial uniform (a delta function)
distribution of money will soon disappear. Let us ask what the eventual
steady state distribution of money after many such moves will be?
At each move, a pair of randomly chosen players share a random fraction
of their total money among themselves.
The answer is well established in physics for more than a century --- soon,
there will be a stable money distribution and it will be Gibbs distribution:
$P(m) \sim \exp[-m/T]$; $T=M/N$ \cite{cc:Dragulescu:2000}.
(b) Now think of a modified move in this game: each player `saves' a fraction
$\lambda$ of the money available with him/her after each move and while going
to the next move. Everybody saves the same fraction $\lambda$.
What is the steady state
distribution of money after a large number of such moves?
It becomes Gamma-function like, while the distribution parameters
of course depend on $\lambda$
(see Ref.~\cite{cc:Chakraborti:2000,cc:Patriarca:2004})
see also Ref.~\cite{cc:Angle:1986,cc:Angle:2006};
for a somewhat different model with similar results developed much earlier.
No exact treatment of this problem is available so far.
(c) What happens to the eventual money distribution among these players if
$\lambda$ is not the same for all players but is different for different
players? Let the distribution $\rho(\lambda)$ of saving propensity
$\lambda$ be such
that $\rho(\lambda)$ is non-vanishing when $\lambda \to 1$.
The actual distribution will depend on the saving propensity distribution
$\rho(\lambda)$, but for all of them, the asymptotic form of the
distribution will become
Pareto-like: $P(m) \sim m^{-\alpha}$; $\alpha=2$ for $m \to \infty$.
This is valid for all such distributions (unless $\rho(\lambda) \propto
(1-\lambda)^\beta$, when $P(m) \sim m^{-(2+\beta)}$).
However, for variation of $\rho(\lambda)$ such that $\rho(\lambda) \to 0$
for $\lambda < \lambda_0$, one will get an initial Gamma function form
for $P(m)$ for small and intermediate values of $m$, with parameters
determined by $\lambda_0$ ($\ne 0$), and this distribution will
eventually become Pareto-like for 
$m \to \infty$~\cite{cc:Chatterjee:2003,cc:Chatterjee:2004,cc:Repetowicz:2005}.
A somewhat rigorous analytical 
treatment of this problem is now available~\cite{cc:Mohanty:2006}.

A major limitation of these money-only exchange models 
considered earlier~\cite{cc:EWD05,cc:othermodels,cc:marjit,cc:Dragulescu:2000,cc:Chakraborti:2000,cc:Hayes:2002,cc:Chatterjee:2004,cc:Chatterjee:2003,cc:Chakrabarti:2004,cc:Slanina:2004,cc:ESTP:KG,cc:EWD:CC,cc:Angle:1986,cc:Angle:2006,cc:Patriarca:2004,cc:Repetowicz:2005,cc:Mohanty:2006,cc:Chatterjee:2005,cc:APFA5}
(and summarised in (a), (b) and (c) above) is that
it does not make any explicit reference to the commodities exchanged with
the money and to the constraints they impose. 
Also, the wealth is not just the money is possession (unless the commodity
exchanged with the money is strictly consumable).
Here, we study the effect
of a single non-consumable commodity on the money (and also wealth)
distributions in the steady state, allowing for the total (in time)
price fluctuation.
This allowance of price fluctuation here is very crucial for the model;
it allows for the stochastic dynamics to play its proper role
in the market.
As such, this model is therefore quite different from that considered
recently in Ref.~\cite{cc:Ausloos:2006}, where $p_0$ is strictly unity
and the stochasticity enters from other exogenous factors.
In the sense that we also consider two exchangeable variables in the market,
our model has some similarity with that in Ref~\cite{Silver:2002}.
However, Silver et al~\cite{Silver:2002}
consider only random exchanges between them (keeping the total conserved)
while we consider random exchanges permitting price fluctuations
and savings. As such they get only
the Gamma distribution in wealth, while we get both Gamma and Pareto
distributions.

In spite of many significant effects due to the inclusion
of a non-consumable commodity,
the general feature of Gamma-like form of the money (and wealth)
distributions (for uniform $\lambda$) and the power law tails
for both money and wealth (for distributed $\lambda$) 
with identical exponents, are seen to
remain unchanged. The precise studies (theories) for the money-only
exchange models are therefore extremely useful and relevant.

Specifically, we study here numerically the behavior of an ideal gas 
like model of markets having only one non-consumable commodity. 
The total amount of money in the market $M=\sum_i m_i$, $i=1,\ldots,N$
is fixed, so is the total amount of commodity $C=\sum_i c_i$ and of course
the total number of agents $N$ in the market. As in the market there is only
one commodity, which is non-consumable, we normalize its global price $p_0= M/C$
to unity. The wealth of any agent $i$ at any time $t$ is therefore
$w_i(t)= m_i(t) + c_i(t)$. If no fluctuation in price $p$ is allowed
(over $p_0$), then the money-commodity exchange leads to trivial money
and commodity distribution as shown in Fig.~\ref{fig:commogif.flam.del0},
which keeps the wealth of any agent unchanged over time. If we now allow
the price $p(t)$ at any time to fluctuate over $p_0$ by a factor $\delta$
(as in (\ref{pricedelta})), nontrivial money, commodity and wealth
distributions set in (the steady states of which are independent of 
$\delta$; $\delta \ne 0$). 
We investigated here the behavior of the 
steady-state distributions of money, commodity and total wealth,
as this dynamics of trading or exchange of money or commodity proceeds,
allowing for temporal fluctuations in the price of the commodity.
These distributions are studied in markets with agents having uniform
(see Figs.~\ref{fig:commogif.flam.mc} and \ref{fig:commogif.flam.w}) 
and random saving factors 
(see Figs.~\ref{fig:commogif.dlam.mc} and \ref{fig:commogif.dlam.w}). 
The self-organizing features in money distribution 
are similar to the cases without any commodity (or with consumable 
commodities), the commodity distribution shows an exponential decay. 
The wealth distribution shows interesting behavior: Gamma like
distribution for uniform saving propensity and has a power-law tail 
(with Pareto exponent value $\nu=1$) for 
a market with agents having random saving propensity.
Although our results are numerical, and the Pareto behavior
for the wealth distribution tail gets somewhat more
restricted in range (compared to that of the money distribution;
see Fig.~(\ref{fig:commogif.dlam.mc})), the robustness of the
power-law behavior nevertheless becomes obvious from, say
Fig.~(\ref{fig:commogif.dlam.mc}), where the power law tail for
the money distribution clearly dominates over the commodity
distribution tail, which rapidly decays off exponentially.
%%%%%%%%%%%%%%%%%%%%%%%%%%%%%%%%%%%%%%%%%%%%%%%%%%%%%%%%%%%%%%%%5

%\begin{acknowledgments}
%\acknowledgments{
We are extremely grateful to
Anindya-Sundar Chakrabarti for useful suggestions and comments.
%\end{acknowledgments
%}

%%%%%%%%%%%%%%%%%%%%%%%%%%%%%%%%%%%%%%%%%%%%%%%%%%%%%%%%%%%%%%%%

%%%%%%%%%%%%%%%%%%%%%%%%%%%%%%%%%%%%%%%%%%%%%%%%%%%%%%%%%%%%%%%%
\end{document}